\documentclass[osa,twocolumn,showpacs,floatfix]{revtex4}

\usepackage{here}

\usepackage{graphicx}

\newcommand\pictc[5]{\begin{figure}
                       \centerline{
                       \includegraphics[width=#1\columnwidth]{#3}}
                   \protect\caption{\protect\label{fig:#4} #5}
                    \end{figure}            }
\newcommand\pict[4][1.]{\pictc{#1}{!tb}{#2}{#3}{#4}}
\newcommand\rpict[1]{\ref{fig:#1}}

\newcommand\leqt[1]{\protect\label{eq:#1}}
\newcommand\reqtn[1]{\ref{eq:#1}}
\newcommand\reqt[1]{(\reqtn{#1})}

\newcounter{Fig}

\begin{document}
\begin{sloppy}

\title{Generation and stability of discrete gap solitons}

\author{Andrey A. Sukhorukov}
\author{Yuri S. Kivshar}

\affiliation{Nonlinear Physics Group, Research School of Physical
Sciences and Engineering, Australian National University,
Canberra, ACT 0200, Australia}

\begin{abstract}
We analyze stability and generation of discrete gap solitons in
weakly coupled optical waveguides. We demonstrate how both stable
and unstable solitons can be observed experimentally in the
engineered binary waveguide arrays, and also reveal a connection
between the gap-soliton instabilities and limitations on the
mutual beam focusing in periodic photonic structures.
\end{abstract}

\ocis{190.4420, 190.5940 }

\maketitle

Suppression of the diffraction-induced beam spreading and
generation of spatial optical solitons is usually associated with
the transverse confinement of a beam in a self-induced waveguide
\cite{Kivshar:2003:OpticalSolitons}. As the medium refractive index increases through
the nonlinear response, the beam becomes self-trapped in the
effective high-index region due to the total internal reflection.
In periodically modulated photonic structures such as fiber Bragg
gratings and arrays of weakly coupled optical waveguides,
nonlinearity-induced localization of light is also possible in the
band-gaps which appear due to the Bragg reflection. The later
mechanism is responsible for the formation of temporal Bragg
solitons~\cite{deSterke:1994-203:ProgressOptics} and spatial gap
solitons~\cite{Mandelik:2003-53902:PRL}.

Temporal Bragg solitons are usually analyzed in the framework of
the coupled-mode theory, where the total field is presented as a
superposition of nonlinearly coupled counter-propagating waves
which experience Bragg reflection. This case corresponds to a
narrow gap in the transmission spectrum, and it is well studied in
the theory of fiber Bragg
gratings~\cite{deSterke:1994-203:ProgressOptics}.

However, the coupled-mode theory accounts for an isolated gap
only, and such a simplified description of the properties of
periodic structures is not valid for analyzing deeper modulations
of the refractive index such as, for example, those in waveguide
arrays and photonic crystals. Deeper gratings demonstrate more
complicated dynamics, including instabilities of nonlinear guided modes
due to multi-gap coupling effects and a resonant interaction
between different bands~\cite{Sukhorukov:2001-83901:PRL}.

In this Letter, we study the stability properties of discrete gap
solitons in engineered binary arrays of optical waveguides~\cite{Sukhorukov:2002-2112:OL}, which have many common properties with spatial gap solitons observed experimentally\cite{Mandelik:2003-53902:PRL}. Our analysis, based on effective discrete equations, captures the main features of discrete solitons, and these results are confirmed by numerical simulations. In particular, we demonstrate how both stable and unstable gap solitons can be observed experimentally, and also reveal a fundamental link between soliton instabilities and limitations on the mutual beam focusing.

We consider the head-on excitation geometry, when optical beams
are incident at grazing angles on a one-dimensional periodic
structure of weakly coupled optical waveguides. Then, the beam
self-action and interaction resulting in soliton formation can be
described by the scalar nonlinear Schr\"odinger equation,
\begin{equation} \leqt{nls}
  i \frac{\partial {\cal E}}{\partial z}
  + \frac{\partial^2 {\cal E}}{\partial x^2}
  + \nu(x) {\cal E}
  + |{\cal E}|^2 {\cal E}
  = 0,
\end{equation}
where $x$ and $z$ are the transverse and longitudinal coordinates,
respectively, ${\cal E}(x,z)$ is the normalized electric field
envelopes, $\nu(x)=\nu(x+h)$ is the normalized refractive index
profile, $h$ is the spatial period, and medium has a self-focusing nonlinear response.

Recently, it has been suggested that the properties of spatial gap
solitons can be engineered in {\em binary waveguide arrays}, where
the first gap is controlled by the width difference between
alternating ``thick'' and ``thin''
waveguides~\cite{Sukhorukov:2002-2112:OL}. Wave localization and
soliton formation in this gap can be described within the
framework of a discrete model of the tight-binding approximation,
where the total field is decomposed into a superposition of weakly
overlapping modes $\varphi_n(x)$ of the individual waveguides in
the form, ${\cal E}(x,z) = \sum_n E_n(z) \varphi_n(x) \exp(i
\lambda_n z)$. After substituting this expression into Eq.~\reqt{nls}, we can obtain the modified discrete nonlinear Schr\"odinger (DNLS) equation  for the normalized mode amplitudes $E_n$,
\begin{equation} \leqt{DNLS}
   i \frac{d E_n}{dz}
   + \lambda_n E_n
   + (E_{n-1} + E_{n+1})
   + \gamma_n |E_n|^2 E_n
   = 0.
\end{equation}
Here $\lambda_n$ characterizes the linear propagation constant of
the mode guided by the $n-$th waveguide, $\gamma_n$ are the
effective nonlinear coefficients, and $\lambda_{2 n + 1} \equiv -
\rho$, $\lambda_{2 n} \equiv 0$, where we assume appropriate
normalization.

Soliton solutions of Eq.~\reqt{DNLS} have the form $E_n = u_n
\exp(i \beta z)$, where $\beta$ is the propagation constant, and
$u_n$ is the soliton profile. Discrete bright solitons should have
exponentially decaying tails, $u_{\pm|n+2 m|} \simeq u_{\pm|n|} \exp( - |2 m| \kappa)$ as $n,m\rightarrow +\infty$. This condition is satisfied inside the band-gaps where ${\rm Re}\kappa \ne 0$. It can be
demonstrated~\cite{Sukhorukov:2002-2112:OL} that $\kappa = (1/2)
{\rm cosh}^{-1}(-\eta/2)$, where $\eta = 2 - \beta (\beta+\rho)$.
Then, discrete solitons can appear in the semi-infinite total
internal reflection gap, $ \beta > -(\rho/2) +
\sqrt{(\rho/2)^2+4}$, and {\em gap solitons} can form inside the
{\em Bragg reflection gap} for $-\rho<\beta<0$.

We can find an analytic estimate for the width of the gap soliton,
a key parameter which can be controlled directly in experiment. In
the total internal reflection gap, the width of the conventional
discrete solitons can vanish for $\beta \rightarrow +\infty$,
until almost all the energy is confined in a single
waveguide~\cite{Bang:1994-205:NLN}. On the other hand, the
situation is qualitatively different for discrete gap solitons,
where the soliton width {\em is bounded from below}, and it can be
estimated according to the linear tail asymptotic,
\begin{equation} \leqt{width}
   \Delta > 4 / \max_{\beta, -\rho<\beta<0} {\rm Re}[\kappa(\beta)]
      = 4 /  {\rm cosh}^{-1}(1 + \rho^2 / 8).
\end{equation}
Thus, the minimum soliton width $\Delta$ is determined by the gap
width ($\rho$), and the strongest localization is achieved at the
middle of the gap ($\beta \sim -\rho/2$). For narrow gaps we have
$\Delta > 8/ \rho \gg 1$, and in this case the soliton dynamics
can be described by the coupled-mode
equations~\cite{deSterke:1994-203:ProgressOptics}. However, for
larger bandgaps the soliton can be localized at several waveguides
($\Delta \simeq 1$), and discreteness effects become very
important.
\pict{fig01.eps}{power2}{ Top: soliton power vs. propagation
constant for A-type (dashed gray) and B-type (black) discrete gap
solitons: solid~--- stable, dashed~--- unstable, and dotted~---
oscillatory unstable. Bottom: examples of the soliton profiles of
(a) A and (b) B symmetries at $\beta = -1$; filled circles and
squares mark the field amplitudes at the narrow and wide
waveguides, respectively; $\rho=1.5$. }

The soliton profiles and associated powers $P = \sum_n |u_n|^2$
are calculated numerically. We find that there exist {\em two
families of gap solitons}, A and B, which bifurcate from the lower
gap edge ($\beta = -\rho$) in a self-focusing medium, see the
example in Fig.~\rpict{power2}. As the power increases, the
soliton width decreases, and it reaches the minimum value inside
the gap, in agreement with the analytic estimate. Profiles of
strongly localized discrete gap solitons are shown in
Fig.~\rpict{power2}(bottom).

\pict{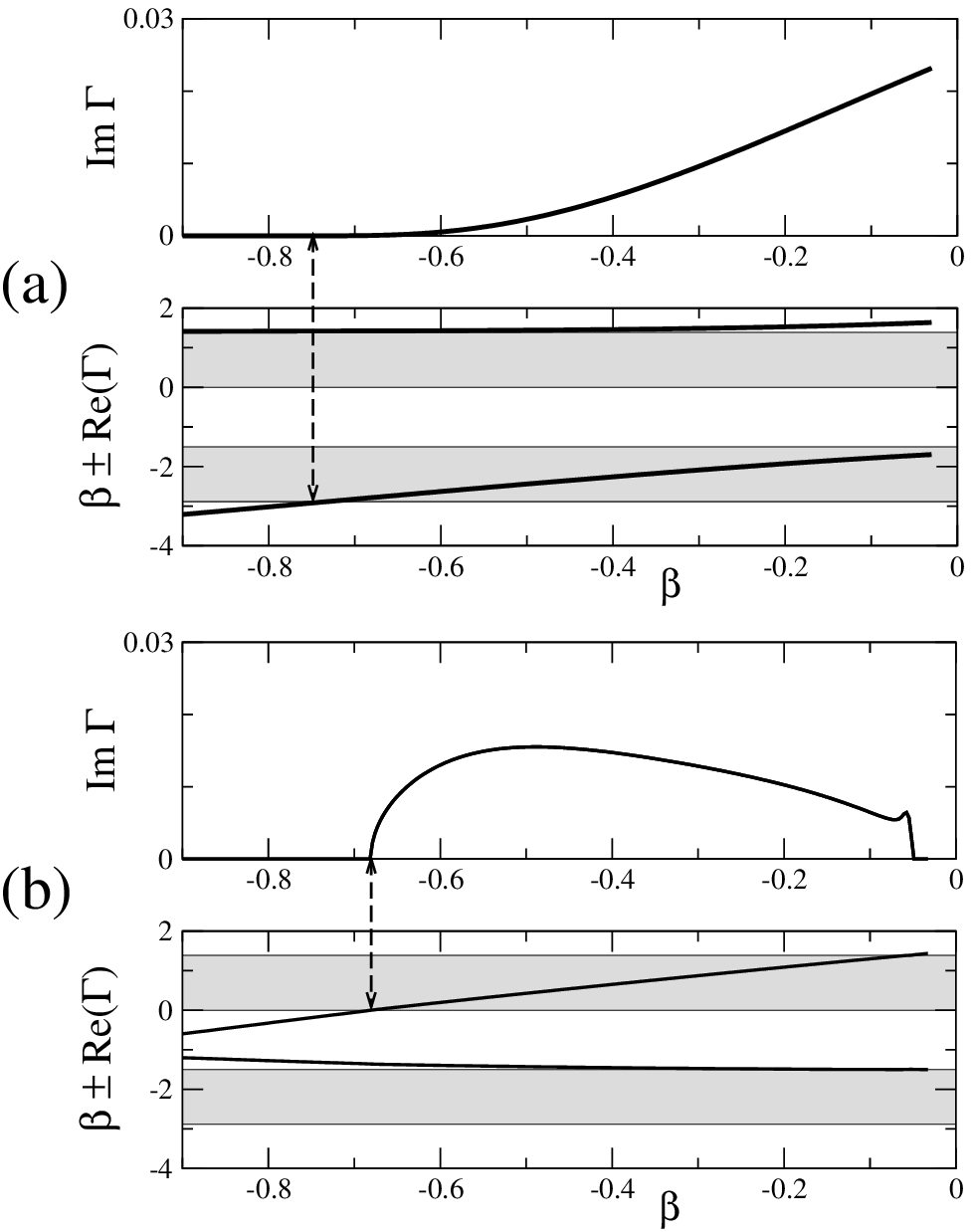}{instab}{ Example of (a)~external and (b)~internal
resonances between an eigenmode of the linear eigenvalue problem
and a bang-gap edges of the continuous spectrum that lead to an
oscillatory instability of the odd-symmetry gap solitons. }

A practically important question is {\em soliton stability}. We
present the field in the form
\begin{equation} \leqt{perturb}
   E_n = (u_n + v_n e^{-i \Gamma z} + w_n^\ast e^{i \Gamma^\ast z})
         e^{i \beta z} ,
\end{equation}
and perform the linear stability analysis for the evolution of
small-amplitude perturbations ($v_n$ and $w_n$) on top of the
soliton profile $u_n$. After substituting Eq.~\reqt{perturb} into
Eq.~\reqt{DNLS}, we obtain a set of linear equations for $v_n$ and
$w_n$ which possess localized solutions at discrete values of
$\Gamma$. Solutions with ${\rm Im}\Gamma > 0$ indicate the soliton
instability, as the perturbations grow exponentially.

We find that the gap solitons of the A-type exhibit symmetry-breaking instabilities, and the instability growth rate becomes
larger for strongly localized modes. This property is common for
the systems with a broken translational symmetry, and it was
demonstrated experimentally that such instabilities can strongly
affect the soliton generation, and they can be utilized for the
soliton steering~\cite{Morandotti:1999-2726:PRL}.

Conventional discrete solitons of ``odd'' symmetry are always
stable~\cite{Morandotti:1999-2726:PRL}. Similarly, we find that
the B-type discrete gap solitons do not exhibit symmetry-breaking instabilities.
However, our analysis reveals that these gap solitons become
unstable {\em above a critical power} due to two types of
the oscillatory instabilities.  First, instability can appear
through an {\em external resonance} with the mode in a different
gap, as illustrated in Fig.~\rpict{instab}(a). A similar
instability mechanism was  discussed earlier for nonlinear defect
modes in a layered medium~\cite{Sukhorukov:2001-83901:PRL}.
Second, the gap soliton can lose stability due to an {\em internal
resonance} within the gap, see Fig.~\rpict{instab}(b). The latter
scenario was first discovered for temporal Bragg solitons~\cite{Barashenkov:1998-5117:PRL}. In both cases,
instabilities are related to {\em resonant excitation of the band
modes}, and the periodic beating is indicated by a non-zero real
part of the instability eigenvalue $\Gamma$.

\pict{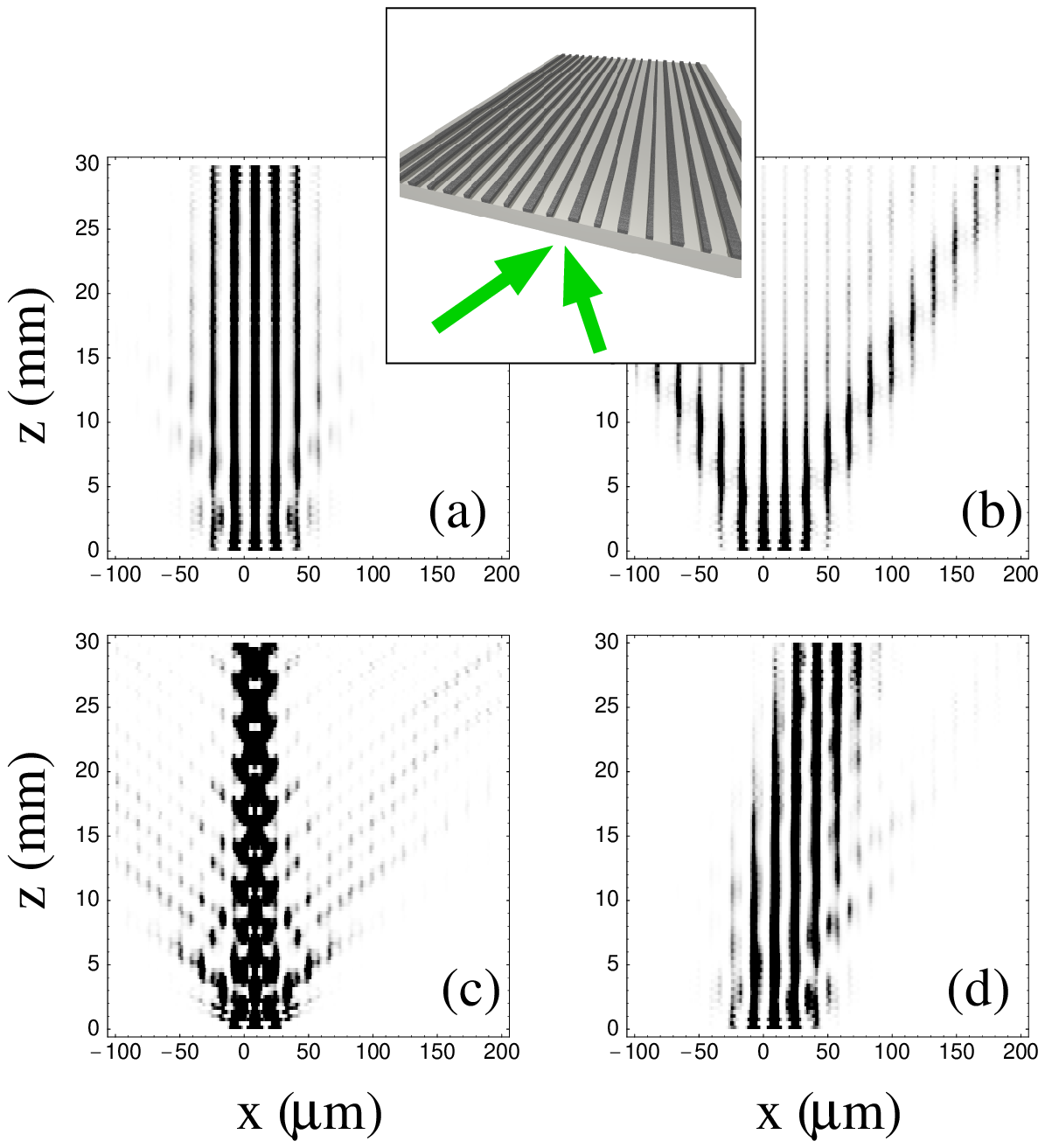}{exciteInterf}{ 
Excitation of discrete
gap solitons by two interfering Gaussian beams: (a)~self-focusing
and soliton formation, (b) self-defocusing when the interference
maxima are located at narrow and wide waveguides, respectively;
(c)~soliton instability at high powers; and (d)~soliton steering
due to power imbalance. 
Parameters are $d_1 = 4 \mu m$, $d_2 = 2.5 \mu m$, $d_s = 5 \mu m$, refractive index $n_0 = 3.4$, effective modulation $\Delta n_0 = 0.0014$, vacuum wavelength $1.5 \mu m$. Insert shows the generation geometry.
}

Soliton excitation inside the frequency gaps of the fiber Bragg
gratings is always accompanied by reflection of a substantial
energy fraction of an incident
pulse~\cite{deSterke:1994-203:ProgressOptics}. The same limitation
is known for the {\em side-on} excitation of spatial gap solitons:
an input beam is almost completely reflected when inclination angle
is tuned deep inside the gap, and the soliton formation is observed
only in the vicinity of gaps~\cite{Mandelik:2003-53902:PRL}. More efficient energy coupling into spatial gap solitons can be
achieved by a {\em head-on} excitation, since the backward
scattering is negligible.

Spatial gap solitons can be generated by two Gaussian beams, which
are tuned to the Bragg resonance and have opposite inclination
angles~\cite{Feng:1993-1302:OL}, as schematically illustrated in Fig.~\rpict{exciteInterf}(insert). If the incident beams are identical, the input field can be written as
\[
   \psi_0(x) =  C e^{ - (x-x_e)^2 / d^2 }
            \cos\left[ \pi (x - x_s) / d \right]
            e^{i (\varphi_1 + \varphi_2)/2}.
\]
where $x_e$ is the beam center, $d$ is width, $C$ is amplitude,
and $\varphi_{1,2}$ are the phases of interfering beams. Parameter
$x_s = (\varphi_2 - \varphi_1) d /2 \pi$ defines a shift of the
interference pattern due to the relative phase between
the beams.

Our analysis of the gap soliton solutions revealed that the
soliton energy is mainly localized at the narrow waveguides, see
Fig.~\rpict{power2}(bottom). We perform numerical simulations
using Eq.~\reqt{nls} and confirm that, when the
interference maxima are at the narrow waveguides, the two Gaussian
beams experience mutual focusing and a gap soliton forms, see
Fig.~\rpict{exciteInterf}(a). The optimal input powers and beam
widths can be chosen to  reduce significantly the emission of
radiation. On the other hand, if the interference maxima appear at
wide waveguides, nonlinear beam interaction accelerates the beam
spreading and break-up [Fig.~\rpict{exciteInterf}(b)]. The soliton
generation by two beams becomes impossible for higher powers [cf.
Figs.~\rpict{exciteInterf}(a) and~\rpict{exciteInterf}(c)], and the generated beam
exhibits periodic beating during its propagation along the
array. This happens due to the development of soliton
instability which limits mutual beam focusing through
a resonant excitation of modes in
different bands, in agreement with our linear stability analysis.
Finally, we illustrate how the soliton motion can be induced by
varying the power imbalance of the input beams
[Fig.~\rpict{exciteInterf}(d)]. 

In conclusion, we have analyzed the basic properties of discrete
gap solitons using engineered binary arrays as an
example. For the first time to our knowledge, we have described
the families of discrete gap solitons and analyzed their
stability. We have revealed two basic mechanisms of the soliton
instability and discussed its connection with mutual beam focusing and gap soliton generation in periodic
photonic structures.

The work was supported by the Australian Research Council.

\end{sloppy}
\end{document}